\documentclass[a4paper,fleqn,usenatbib]{mnras}

\usepackage{newtxtext,newtxmath}
\usepackage[T1]{fontenc}
\usepackage{ae,aecompl}

\usepackage{graphicx}	

\newcommand{\fl}{\emph{Fermi}-LAT~}

\usepackage{multirow}

\title[High energy $\gamma$-ray detection of SNRs in the Large Magellanic Cloud]
{High energy $\gamma$-ray detection of supernova remnants in the Large Magellanic Cloud} 

\author[Campana, Massaro, Bocchino, Miceli, Orlando and Tramacere]  
 {R. Campana$^{1}$$\thanks{E-mail: \url{riccardo.campana@inaf.it}}$,
  E. Massaro$^{2}$,
  F. Bocchino$^{3}$, 
  M. Miceli$^{3,4}$,
  S. Orlando$^{3}$,
  A. Tramacere$^{5}$\\
 $^1$ INAF/OAS, via Gobetti 101, 40129, Bologna, Italy \\
 $^2$ INAF/IAPS, via del Fosso del Cavaliere 100, I-00113 Roma, Italy \\
 $^3$ INAF/Osservatorio Astronomico di Palermo, Piazza del Parlamento 1, I-90134 Palermo, Italy\\
 $^4$ Dipartimento di Fisica e Chimica E. Segr\`e, Universit\`a di Palermo, Piazza del Parlamento 1, 90134, Palermo, Italy\\
 $^5$ Department of Astronomy, University of Geneva, Chemin Pegasi 51, 1290, Versoix, Switzerland
}
      
\date{ Accepted 2022 July 1. Received 2022 June 13; in original form 2022 April 22.}

\pubyear{2022}

\begin{document}

\label{firstpage}
\pagerange{\pageref{firstpage}--\pageref{lastpage}}
\maketitle

\begin{abstract}
We present the results of a cluster search in the $\gamma$-ray sky images
of the Large Magellanic Cloud (LMC) region by means of the Minimum Spanning Tree (MST) 
and DBSCAN
algorithms, at energies higher than 6 and 10 GeV, using 12 years of \fl data. 
Several significant clusters were found, the majority of which associated
with previously known $\gamma$-ray sources.
We confirm our previous detection of the Supernova Remnants N~49B and
N~63A and found new significant clusters associated with the SNRs N~49, N~186D and N~44.
These sources are among the brightest X-ray remnants
in the LMC and corresponds to core-collapse supernovae 
interacting with dense HII regions, indicating that an hadronic origin of high energy photons is the most
likely process.
\end{abstract}

\begin{keywords} $\gamma$-rays: observations --- $\gamma$-rays: source detection ---
Supernova Remnants --- Large Magellanic Cloud.
\end{keywords}

\section{Introduction} 
\label{s:intro}

There is a large consensus in the literature around the hypothesis that Galactic cosmic 
rays (CRs) may originate at shock fronts of expanding supernova remnants (SNRs) via diffusive 
shock acceleration of particles (e.g., \citealt{2005JPhG...31R..95H} and references therein). 
According to this scenario, about $10$\% of the explosion energy of the parent supernova is 
converted in kinetic energy of accelerated CRs with a broad power law energy spectrum, which 
extends up to the PeV domain, where the observed spectrum of CRs in our Galaxy
is characterized by the ``knee''. 
In the last decades, the SNR paradigm for the origin of CRs was successful in reproducing 
the main features of their observed intensity, spectrum 
and chemical composition (e.g., \citealt{2005A&A...429..755P}). 
Furthermore, high energy observations 
have shown that electrons can be accelerated up to TeV energies 
(\citealt{1995Natur.378..255K}) and have allowed to identify the signature of magnetic field amplification 
(\citealt{2003ApJ...589..827B, 2003ApJ...584..758V}), 
an indirect evidence of hadron acceleration at SNR shock fronts (\citealt{2004MNRAS.353..550B}).

However, despite some undisputable successes of the SNR paradigm in reproducing the 
observations, some important issues still remain unsolved. 
$\gamma$-ray observations 
provided 
evidence of accelerated hadrons up to multi-GeV energies 
at SNR shocks (e.g., \citealt{2013Sci...339..807A}) but, to date, there is no convincing observational 
support for the acceleration of hadrons to PeV energies. 
It is even questioned whether SNRs can accelerate hadrons to PeV energies: the observed 
expansion rate of some historical young SNRs (Cassiopeia A, Tycho and Kepler) is too low 
to allow the acceleration of CRs to the energy domain of the knee \citep{2013MNRAS.431..415B}. 

To shed some light on the above issues, we need to analyze carefully the spectrum 
of Galactic CRs above GeV energies and to provide physical insight on the origin of the 
features observed. 
Thus, $\gamma$-ray observations of SNRs are the most natural tool to challenge the SNR 
paradigm and obtain observational evidence in favour or against this scenario 
\citep{2017AIPC.1792b0002G}. 
A first important step is to enlarge the sample of $\gamma$-ray sources associated with 
SNRs in order to improve the statistics of possible CRs accelerators.

The Large Magellanic Cloud (LMC) is a nearby small galaxy rich of young astronomical 
objects and Supernova Remnants (SNR).
It was proposed as an interesting target in the early period of $\gamma$-ray astronomy 
by \citet{ginzburg72}, subsequently developed by \citet{ginzburg84}, for testing the 
metagalactic origin of cosmic rays via the $\pi^0$ decay process.   
The discovery of the emission from LMC at energies higher than 100 MeV was reported by 
\citet{sreekumar92}, on the basis of EGRET-CGRO observations, who evaluated that the cosmic rays in the LMC have a density comparable to the ones in our Galaxy.
Several compact high energy sources were found in the LMC, since the discovery on 
1979 March 5 of the first magnetar SGR 0526$-$66 \citep{mazets79a,mazets79b} later 
associated with the SNR N~49 \citep{cline82}.
\citet{abdo10} reported a first study of the LMC region based on the 11-month 
\emph{Fermi}-Large Area Telescope (LAT) observations and revealed the massive star 
forming region 30 Doradus as a bright high energy source.
\citet{ackermann16} performed the first detailed analysis of the LMC region considering a 
much richer \fl data spanning the first 6 years of the mission and detected four sources.
In the last version (DR2) of the 4FGL catalogue released by the \fl collaboration
\citep[]{abdollahi20,ballet20} in the LMC region there are 18 sources, 
detected in the 50 MeV -- 1 TeV energy range, of which 4 are reported as
extended features and 7 are present also in the 3FHL catalogue \citep{ajello17}.
Furthermore, possible background extragalactic counterparts (unclassified blazar or BL Lac
objects) are indicated for 9 of these 4FGL sources.

At TeV energies, the  H.E.S.S. collaboration observed three sources up to about 10~TeV 
in the LMC \citep{abramowski15}, two of them associated with the SNRs N~157B 
\citep{abramowski12} and N~132D, the former containing the highest known spin-down 
luminosity pulsar PSR J0537$-$6910 \citep{marshall98,cusumano98} with the fast period 
of 16 ms.

\citet{campana18}, hereafter Paper I, used the data collected by the \fl telescope 
\citep{ackermann12} in 9 years of operation for a cluster search at energies higher 
than 10 GeV, and reported the discovery of high energy emission from the two SNRs 
N~49B and N~63A.
In the present paper we extend this analysis of the LMC region using a richer data set 
of 12 years of \fl observation.
As in Paper I we applied the Minimum Spanning Tree (hereafter MST, 
\citealt{campana08,campana13}) source-detection method for extracting photon clusters 
likely corresponding to genuine $\gamma$-ray sources, as already successfully applied 
in a series of papers \citep{bernieri13,paperI,paperII,paperIII,paperIV,campana17,
campana18b,campana21}.
Furthermore, the analysis was verified by an independent cluster search using the DBSCAN algorithm \citep{2013A&A...549A.138T}.

The results of this new analysis confirm the previous SNR detections with a higher 
significance, and found some other photon clusters at positions very close to those of 
SNRs not previously associated with $\gamma$-ray emitters.

The outline of this paper is as follows. 
In Section~\ref{s:selmst} the data selection and the MST and DBSCAN algorithms are described, 
while the analysis of the LMC region is presented in Section~\ref{s:analysis}. 
In Section~\ref{s:snr} we deepen the analysis of some interesting SNR counterparts to 
MST clusters, while in Section~\ref{s:conclusions} the results are summarized.

\section{Photon selection and cluster analysis of LMC high energy $\gamma$-ray data}   
\label{s:selmst}

\subsection{Data selection in the LMC region}    
\label{ss:dataLMC}

\begin{figure*}  
\centering
\includegraphics[width=0.75\textwidth]{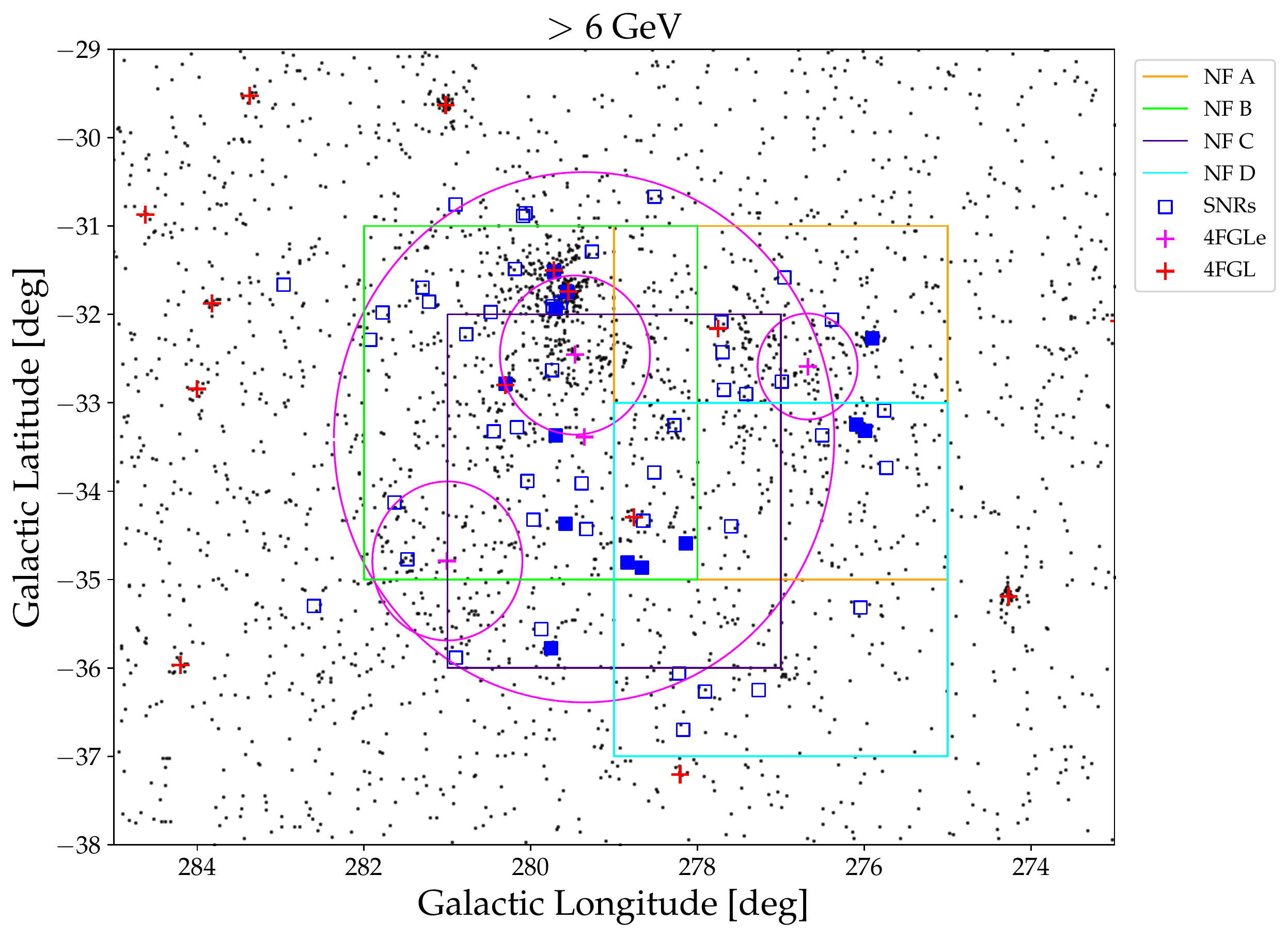}
\includegraphics[width=0.75\textwidth]{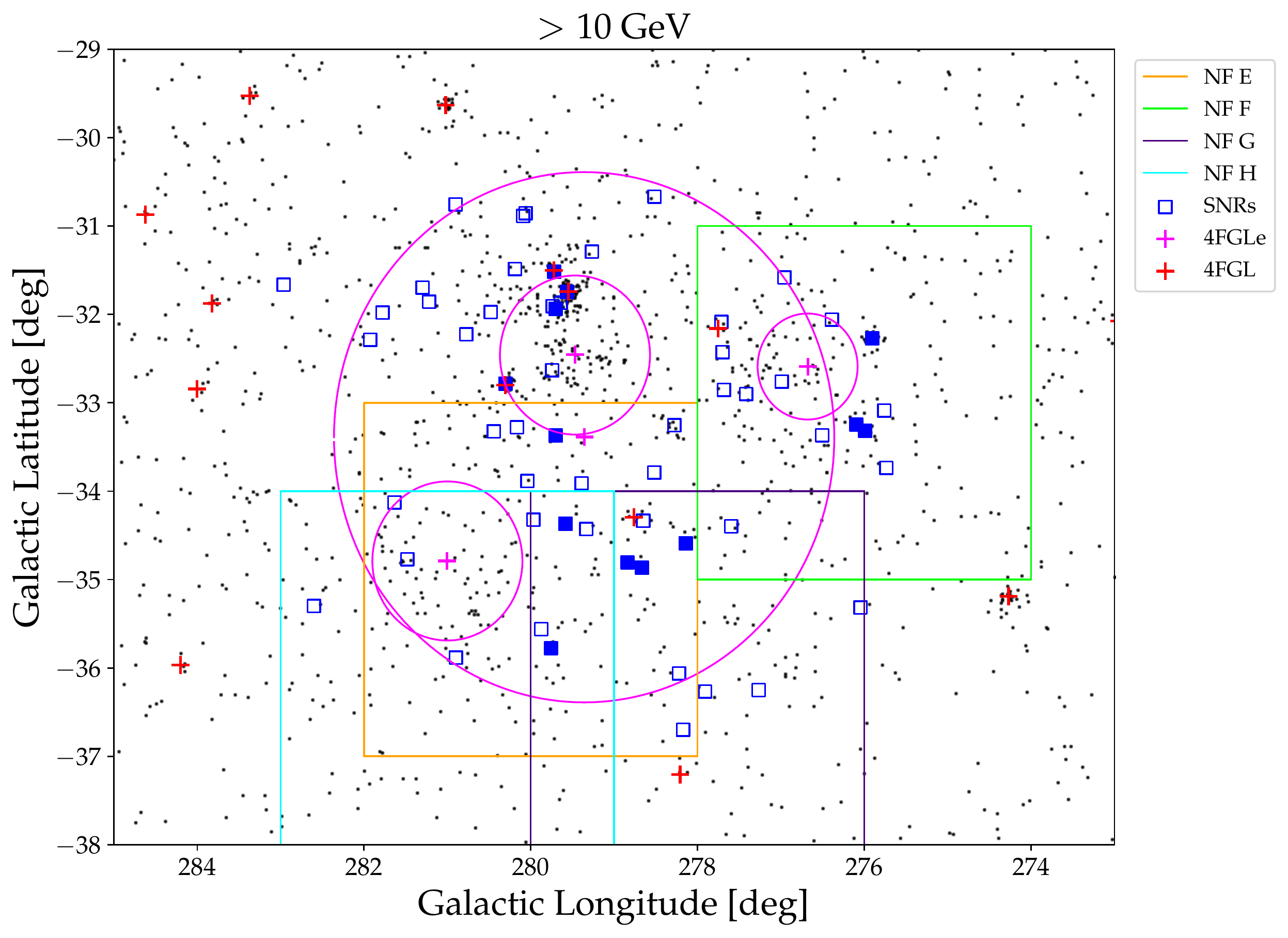}
\caption{Photon maps in equatorial coordinates of the sky region 
(12\degr$\times$9\degr\ in a Galactic coordinate frame) centered at the LMC, at energies 
higher than 6~GeV (upper panel) and higher than 10~GeV (lower panel).
Black points in each panel are the photons.
Magenta circles (centred at the magenta crosses) circles correspond to the sizes of 
the extended 4FGL-DR2 sources, while red crosses mark the other 4FGL sources.
Blue squares are the SNRs in the \citet{maggi16}  catalogue: those marked by 
a filled square have with a X-ray luminosity higher than 10$^{36}$ erg/s.
Large square regions, shown for clarity's sake split in the two panels, mark the narrow 
fields (NF) considered in our analysis: note that only NF B in the upper panel includes 
the bright 30 Doradus complex.
}
\label{skymap_f1}
\end{figure*}

The full \fl dataset, including events above 3 GeV in the 12 years from 
2008 August 4 to 2020 August 4, and processed with the Pass 8, release 3, reconstruction 
algorithm and responses, was downloaded in the form of weekly files from the FSSC 
archive\footnote{\url{http://fermi.gsfc.nasa.gov/ssc/data/access/}}. 
The event lists were then filtered applying the standard selection criteria on data 
quality and zenith angle (source class events, \texttt{evclass} 128), front and back 
converting (\texttt{evtype} 3), up to a maximum zenith angle of $90\degr$. 
Events were then screened for standard good time interval selection.
Then, a region $12\degr\times9\degr$, in Galactic coordinates $273\degr < l < 285\degr$
and $-38\degr < b < -29\degr$ and approximately centred at the LMC, was selected 
and the MST algorithm was applied. 
Figure~\ref{skymap_f1} shows the photon maps in this region at energies higher than 6 
GeV (upper panel) and higher than 10 GeV (lower panel).

The general structure of the $\gamma$-ray emission in the LMC region is very complex.
The last DR2 release of the 4FGL catalogue reports several sources as extended and only
6 of them are at Galactic latitude higher than 20\degr: four are in the LMC, one in the 
Small Magellanic Cloud (SMC) and the last one in the Fornax A region.

One of the extended features, 4FGL J0519.9$-$6845e, has a radius of 3\degr\ and therefore 
it covers the entire central region of LMC including the very bright complex of 30 Dor 
and about 75\% of the known SNRs, while the other three have smaller radii and are in 
localised bright regions inside the largest one.
In Figure~\ref{skymap_f1} these regions are plotted as large magenta circles.
The same catalogue reports 13 more point sources within the considered regions: 5 are
in the circle of the largest one while 8 are outside this boundary, and the latter ones 
have possible associations with confirmed or candidate blazars.

Given such conditions of a high local diffuse emission, the sorting of point-like
sources is a quite difficult work, and particular approaches should be adopted.
Our basic choice is to search clusters at energies higher than 10 GeV, since the
diffuse component is rather weak in this band.
Once a list of clusters is obtained, we searched for correspondent features at lower 
energies and, if they were found, the cluster was considered to be confirmed.
However, in these analyses we had to use smaller fields not including the 30 Dor nebula,
because its relatively high number of photons produces a too short mean angular 
separation in the field for a safe cluster search.
In these analyses at lower energies the resulting cluster parameters were generally
much lower than above 6 GeV, and the extraction of clusters was less stable: in 
particular, small changes in the selection parameters could give different structures.

The LMC is particularly rich of SNRs.
The work of \citet{maggi16} reports 59 remnants.
A richer catalogue, which includes the previous one, adding another group of 15 SNR candidates, is provided by \citet{bozzetto17}; a further sample of 19 optically selected SNR candidates
is given by \citet{yew21}.
As it will be clear from the results of this work, we did not find any well established
positional correspondence between clusters and the \citet{yew21} sources and the \citet{bozzetto17} candidate sources, and therefore
we will limit our analysis to the \citet{maggi16} sample, which contains the brightest
and genuine SNRs.

\subsection{Description of photon cluster detection algorithms}    
\label{ss:alg}

\subsubsection{Minimum Spanning Tree}
\label{sss:mst}

The MST (see, for instance, \citealt{cormen09} and also \citealt{campana08,campana13}) 
is a topometric tool that can be used for searching spatial concentrations in a field 
of points.
The application of this method to the $\gamma$-ray sky and detailed descriptions of its theoretical principles and 
selection criteria were presented elsewhere (e.g., in \citealt{campana18}). As a consequence, 
we summarize here only the properties the main parameters used for the cluster selection.

In a two-dimensional set of $N_n$ points (\emph{nodes}), one can construct the unique 
\emph{tree} (a graph without closed loops) which connects all the nodes with the minimum 
total weight, $\min [\Sigma_i \lambda_i]$, where the elements of the set $\{\lambda\}$ 
are, in this case, the angular distances between photon arrival directions. 

The \emph{primary} cluster selection is then performed in two following steps:
\emph{i) separation},  the removal of all the edges having a length 
$\lambda \geq \Lambda_\mathrm{cut}$, usually defined in units of the mean edge 
length $\Lambda_m = (\Sigma_i \lambda_i)/N_n$ in the MST;
\emph{ii) elimination}: removal of all the sub-trees having a number of nodes 
$N \leq N_\mathrm{cut}$. This leaves only the clusters with a number of photons higher
than this threshold. 
In this paper we adopt the simplifying notation $\Lambda_\mathrm{cut} = 0.6$, implying
the factor $\Lambda_m$.
A successive \emph{secondary} selection is applied for extracting the most robust 
candidates for $\gamma$-ray sources. 
The most important parameter here is the cluster \emph{magnitude}:
\begin{equation}
M_k = N_k g_k  
\end{equation}
where $N_k$ is the number of nodes in the cluster $k$ and the \emph{clustering parameter} 
$g_k$ is the ratio between $\Lambda_m$ and $\lambda_{m,k}$, the mean length over the $k$-th 
cluster edges. 
\cite{campana13} found that $\sqrt{M}$ is well correlated with other statistical 
source significance parameters, in particular with Test Significance from the Maximum Likelihood 
\citep{mattox96} analysis. 
A lower threshold value of $M$ around 15--20 would reject the large majority of spurious 
low-significance clusters.

The centroid coordinates are computed by means of a weighted mean of the 
cluster photon arrival directions \citep[see][]{campana13}. The radius of 
the circle centred at the centroid and containing the 50\% of photons in the cluster, 
the \emph{median radius} $R_m$, for a genuine point-like $\gamma$-ray source
should be smaller than or comparable to the 68\% containment radius of instrumental 
Point Spread Function (PSF).
This radius varies from 0\fdg2 at 3 GeV to 0\fdg14 at 10 GeV in the case of a bright 
source for diffuse class front-observed events \citep{ackermann13a}.
It is also be expected that the angular distance between the cluster centroid and 
a possible counterpart would be lower than the latter value.
We also define the \emph{maximum radius} $R_\mathrm{max}$ as the distance between the centroid 
and the farthest photon in the cluster, usually of some arcminutes for a point-like 
source and a few tens of arcminutes either for extended structures or for unresolved 
close pairs of sources.

In the present analysis we adopted the following criteria for the secondary selection.
For clusters having a number of photons $N > 5$, a threshold of $M \ge 20$ was adopted,
while for clusters having $N = 4$ and $N = 5$ nodes, a threshold of $g \ge 3.5$ and 
$g \ge 3.0$ (corresponding to $M\ge14$ and $M\ge15$, respectively) was chosen in order 
to extract only structures with a photon density much higher than the surrounding and reduce 
in this way the probability for random clustering, as suggested by numerical simulations 
\citep{campana13}. 
We also extended our analysis to low significance clusters ($M < 20$), because of the 
possible associations either with extended features, or with faint sources, as verified 
by \emph{a posteriori} comparisons with literature data sets.
According to \citet{campana13} we expect that a percentage of around 50\% of these low
significance and ``poor'' clusters could be spurious and therefore those expected to be
associated with an interesting counterparts must be validated by further analysis.

\subsubsection{Narrow Field analysis}
\label{sss:ver}

We verified the stability of these clusters performing some further targeted 
searches in several suitably selected narrow fields, hereafter NF.
The use of these smaller-sized fields implies a variation of the mean angular 
separation between photons because of the different occurrence of high photon 
density regions. 
We considered eight $4\degr\times4\degr$ fields selected for including some 
subsample of detected clusters, shown in Figure~\ref{skymap_f1}, where the 
fields are identified by a letter from A to H.
These fields were generally selected to exclude the bright complex of 30 Dor, 
with the only exception of NF B, and partially NF C, and therefore they are 
useful for detecting clusters not very rich and dense.
The lower values of the considered energy intervals were 6, 8, 10 and 12 GeV.
A cluster is therefore considered \emph{very stable} (VS) when it is found 
with high $M$ values in all narrow field searches, \emph{stable} (S) when it 
results with a low $M$ only in some selections, and  \emph{poorly stable} (PS) 
when it appears only in a small number of fields generally with a low $M$.

\subsubsection{DBSCAN}
\label{sss:dbs}
The \emph{Density Based Spatial Clustering of Applications with Noise} algorithm (DBSCAN, \citealt{Ester96DBScan}), is a topometric density-based clustering method, that uses local 
density of points to find clusters in data sets that are affected by background noise.
Thanks to its embedded capability to distinguish background noise (even when the background is not uniform),  it has been successfully used to detect sources in $\gamma$-ray  photon lists \cite{2013A&A...549A.138T} and in CCD optical images \cite{2016MNRAS.463.2939T}.
For a detailed description we refer to \cite{2013A&A...549A.138T},
and in the following we give only a brief description of the method and its application.

Let $S$ be a set of photons, where each element is described by the sky coordinates. The distance between two given elements $({p_l,p_k})$ is defined as 
the angular distance on the unit sphere, $\rho(p_l,p_k)$. 
Let $N_{\varepsilon}(p_i)$ be the  set of points contained within a 
radius $\varepsilon$, centred on  $p_i$, and $|N_{\varepsilon}(p_i)|$ the number of contained points, i.e., the estimator 
of the local density, and $K$ a threshold value. Clusters are built according to the 
local density around  each point $p_i$. A  point is  classified according to the local density  defined as:

\begin{itemize}
	\item \textit{core point}: if   $|N_{\varepsilon}(p_i)|\geq K$ .
	\item \textit{border point}: if  $|N_{\varepsilon}(p_i)|< K$, but
	at least one core point belongs to $N_{\varepsilon}(p_i)$. 
	\item \textit{noise point}, if both  the conditions above are not satisfied.
\end{itemize}

Points are classified according to their inter-connection as: 
\begin{itemize}
\item \textit{directly density reachable}:  a point $p_j$  is defined \textit{density reachable} from a point $p_k$, if $p_j \in N_{\varepsilon}(p_k) $ and $p_k$ is a \textit{core} point.
\item \textit{density reachable}: a point $p_j$  is defined \textit{density reachable} from a point $p_k$, if exists a chain of  \textit{directly density reachable} points connecting, 
 $p_j$  to  $p_k$. 
\item \textit{density connected}: two points $p_j$, $p_k$  are defined \textit{density connected}  if exits a \textit{core} point $p_l$ such that both  $p_j$ and $p_k$, are  \textit{density reachable} from $p_l$.
\end{itemize}
The DBSCAN builds the cluster by recursively connecting \textit{density connected}  points to each set of \textit{core} points found in the set. 
Each cluster $C_m$   will be described by  the position of the centroid ($x_c$,$y_c$), the ellipse 
of the cluster containment,  the number of photons in the cluster ($N_p$), and the significance.
The ellipse of the cluster containment  is defined by major and 
minor semi-axis ($\sigma_x$  and $\sigma_y$)  and the inclination angle
($\sigma_{\alpha}$)  { of the major semi-axis w.r.t. the latitudinal coordinate} ($b$ or $DEC$). 
The ellipse is evaluated using the principal component analysis method (PCA, \citealt{Jolliffe1986}). 

The detection algorithm proceeds as follows:
\begin{enumerate}
    \item The initial data set is pre-processed, selecting all the photons within an extraction radius ($R_\mathrm{extr}$), and above a given threshold energy $E_\mathrm{th}$.
    \item The scanning radius of the DBSCAN $\varepsilon$ is set equal to the \fl PSF at $E_\mathrm{th}$.
    \item The background level, $K_\mathrm{bkg}$, is heuristically estimated as described in the following, and the final cluster list $C$ is produced performing a DBSCAN run with $K=K_\mathrm{bkg}$ and $\varepsilon=\mathrm{PSF}(E_\mathrm{th})$
\end{enumerate}

The background estimation algorithm proceeds as follows:
\begin{enumerate}
 \item In a first step, $K$ is estimated from to the average photon density within $R_{\rm extr}$, i.e. $K_\mathrm{ave} = \frac{N_\mathrm{tot}\Omega_{\varepsilon}}{\Omega_{R_\mathrm{extr}}}$, where $\Omega_{R_\mathrm{extr}}$ is the angular size of the extraction region, $\Omega_{\varepsilon}$ is the angular size of the scanning brush, and  $N_\mathrm{tot}$ is the total number of photons within the extraction region.
 \item A first run of the DBSCAN is performed with 
 $K=K_\mathrm{ave}+K_\mathrm{frac} \sqrt{K_\mathrm{ave}}$, producing a cluster list $C^*$.
 \item All the photons belonging to the detected clusters are flagged as signal events, and the remaining as background events. The final value of background density, $K_\mathrm{bkg}$,  is evaluated as the average value of background photons within each circular region centred on each background photon, within a distance equal to $\varepsilon$.
 \item The final detection is performed by setting $K = K_\mathrm{bkg} + K_\mathrm{frac} \sqrt{K_\mathrm{bkg}}$.
\end{enumerate}

The significance of a cluster ($S_\mathrm{cls}$), evaluated  according to the signal-to-noise  Likelihood Ratio Test (LRT) method proposed by \cite{LiMa1983},  and  explained in detail in \cite{2013A&A...549A.138T}, is defined as:
\begin{equation}
       S_{\rm cls}=\sqrt{2 \left  ( N^{\rm in}_{\rm src} \ln \left[ \frac{2 N^{\rm in}_{\rm src}}{N^{\rm in}_{\rm src}+N^{\rm eff}_{\rm bkg}} \right]+
       N^{\rm eff}_{\rm bkg} \ln \left[ \frac{2N^{\rm in}_{\rm src}}{ N^{\rm in}_{\rm src}+N^{\rm eff}_{\rm bkg}} \right ]  \right  ).}
       \label{eq:signif}
\end{equation} 
where $N^\mathrm{in}_\mathrm{src}$ represents the number of  points in each cluster, and $N^\mathrm{eff}_\mathrm{bkg}$ is the expected number of background photons within the circle enclosing the most distant cluster point,  according to the estimated value of $N_\mathrm{bkg}$.
Assuming that a cluster is due to a background fluctuation, the variable $S_\mathrm{cls}^2$ is expected to follow a chi square distribution, with one degree of freedom ($\chi(1)^2$), hence it defines our test statistics (TS). The value of $S_\mathrm{cls}^2$ is highly correlated with the $\sqrt{TS_\mathrm{LAT}}$ returned by the \fl analysis (using the same value of $E_\mathrm{th}$) with:  $S_\mathrm{cls}\approx 0.5 \sqrt{TS_\mathrm{LAT}}$ \citep{2013A&A...549A.138T}.

\begin{table*}
\caption{Coordinates and main properties of MST clusters with $M \ge 20$ detected in the LMC 
sky region at energies higher than 10 GeV (1st section) and higher than 6 GeV (2nd section).  
$\Lambda_\mathrm{cut}$ values are reported in the first line of each part.
Celestial coordinates are J2000, angular distances $\Delta \theta$
are computed between the centroids of MST clusters and those of indicated counterparts.
The letter ``e'' indicates likely extended structures, see main text for details.
Clusters not included in NFs are marked by a dash.
}
\centering
{\small
\begin{tabular}{rrrrrccrrrrl}
\hline
RA~~~ & Dec ~~~& $l$~~~ & $b$~~~ & $N$ & $g$~~~ & $M$~~~ & $R_m$ & $R_\mathrm{max}$ & $\Delta \theta$ & Stability &  Possible \\
\degr~~ & \degr~~  & \degr~~  & \degr~~  &     &        &        & \arcmin~   &  \arcmin~      &   \arcmin~  &  &  counterparts\\
\hline
\hline
\multicolumn{12}{c}{$E>10$ GeV, $\Lambda_\mathrm{cut} =  0\fdg6 = 5\farcm5$}\\
\hline
  77.440 & $-$64.308 & 274.297 & $-$35.206 & 21 &  3.739 &  78.517 &  3.8 & 13.9 &  1.65 & -- &  4FGL J0509.9$-$6417     \\  
  80.630 & $-$67.891 & 278.293 & $-$33.311 &  8 &  2.782 &  22.259 &  4.8 &  8.9 &  3.45 & VS &  N~44                    \\  
  81.284 & $-$69.618 & 280.275 & $-$32.780 & 16 &  3.446 &  55.132 &  4.1 & 15.3 &  1.96 & VS &  4FGL J0524.8$-$6938, P4 \\  
         &           &         &           &    &        &         &      &      &  1.35 &    &  N~132D                  \\
  81.379 & $-$65.928 & 275.919 & $-$33.316 &  6 &  3.862 &  23.175 &  2.8 &  7.7 &  3.66 &  S &  N~49B                   \\  
  81.515 & $-$66.093 & 276.106 & $-$33.239 &  9 &  2.372 &  21.345 &  5.4 &  8.2 &  0.77 &  S &  N~49                    \\  
  81.700 & $-$69.129 & 279.672 & $-$32.719 & 17 &  2.833 &  48.161 &  8.4 & 13.9 &  6.28 & PS &  SNR B0528$-$692  ?      \\  
  82.443 & $-$68.704 & 279.124 & $-$32.521 &  8 &  3.514 &  28.110 &  3.7 &  5.7 & 17.82 & VS &  4FGL J0530.0$-$6900e    \\  
  82.714 & $-$69.199 & 279.688 & $-$32.352 & 11 &  3.419 &  37.610 &  4.9 &  7.0 & 12.80 & VS &  4FGL J0530.0$-$6900e    \\  
  83.939 & $-$69.490 & 279.957 & $-$31.884 & 10 &  2.813 &  28.133 &  5.7 & 11.1 &       & PS &                          \\  
  83.955 & $-$66.051 & 275.923 & $-$32.261 & 12 &  3.161 &  37.933 &  4.6 &  8.1 &  0.94 & VS &  N~63A                   \\  
e 84.278 & $-$69.160 & 279.554 & $-$31.806 & 77 &  4.803 & 369.868 &  7.8 & 24.8 &  3.80 & VS &  4FGL J0537.8$-$6909, P2 \\  
         &           &         &           &    &        &         &      &      &  3.60 &    &  N~157B                  \\
  85.186 & $-$69.329 & 279.707 & $-$31.467 & 15 &  2.924 &  43.860 &  5.2 & 13.7 &  2.15 &  S &  4FGL J0540.3$-$6920, P1 \\  
         &           &         &           &    &        &         &      &      &  2.99 &    &  PSR J0540$-$6919        \\
         &           &         &           &    &        &         &      &      &  2.98 &    &  SNR B0540$-$693         \\
  90.281 & $-$70.568 & 280.996 & $-$29.636 & 19 &  5.449 & 103.534 &  2.6 &  9.2 &  2.08 & -- &  5BZQ J0601$-$7036       \\  
\hline
\hline
\multicolumn{12}{c}{$E>6$ GeV, $\Lambda_\mathrm{cut} =  0\fdg7 = 4\farcm5$}\\
\hline
  74.306 & $-$66.225 & 276.949 & $-$36.071 & 11 &  2.145 &  23.600 &  4.4 & 11.1 &  1.25 & VS &  1WGA~J0457.1$-$6612   \\       
  74.875 & $-$70.169 & 281.536 & $-$34.794 &  7 &  3.142 &  21.994 &  1.9 &  6.4 &  3.10 & VS &  N~186D                \\       
  77.506 & $-$64.303 & 274.285 & $-$35.179 & 31 &  3.021 &  93.653 &  4.6 & 18.8 &  0.96 & -- &  4FGL J0509.9$-$6417   \\  
  77.934 & $-$68.137 & 278.817 & $-$34.254 &  9 &  2.447 &  22.019 &  5.3 &  7.6 &  3.73 & VS &  4FGL J0511.4$-$6804   \\       
         &           &         &           &    &        &         &      &      &  2.37 &    &  PMN J0511$-$6806 bcu  \\  
  78.936 & $-$72.681 & 284.028 & $-$32.871 & 11 &  2.056 &  22.615 &  5.0 &  9.9 &  2.05 & -- &  4FGL J0516.1$-$7240   \\       
         &           &         &           &    &        &         &      &      &       &    &  PKS 0517$-$726 bcu    \\ 
  80.627 & $-$67.933 & 278.344 & $-$33.305 & 18 &  2.102 &  37.830 &  5.9 & 11.0 &  4.49 & VS &  N~44                  \\
  81.199 & $-$69.666 & 280.337 & $-$32.801 & 22 &  2.670 &  58.732 &  6.2 & 16.9 &  1.54 & VS &  4FGL J0524.8$-$6938 P4  \\
         &           &         &           &    &        &         &      &      &  2.11 &    &  N~132D                \\    
  81.427 & $-$65.992 & 275.992 & $-$33.288 & 17 &  2.450 &  41.646 &  4.6 & 11.8 &  5.68/1.80 & VS &  N~49/N~49B       \\
  81.691 & $-$69.016 & 279.540 & $-$32.740 & 13 &  2.176 &  28.285 &  5.0 &  9.7 &       &  S &                        \\       
  82.317 & $-$72.740 & 283.834 & $-$31.878 &  9 &  2.858 &  25.722 &  2.9 &  5.9 &  0.49 & -- &  4FGL J0529.3$-$7243   \\       
         &           &         &           &    &        &         &      &      &       &    &  PKS 0530$-$727 bcu    \\ 
  82.439 & $-$68.993 & 279.464 & $-$32.479 &  8 &  2.575 &  20.601 &  3.3 &  7.4 &  1.36 &  S &  4FGL J0530.0$-$6900e  \\
  82.477 & $-$68.715 & 279.135 & $-$32.507 &  9 &  3.657 &  32.917 &  2.1 &  4.9 &       & VS &                        \\
  82.731 & $-$69.217 & 279.708 & $-$32.343 & 19 &  2.601 &  49.413 &  5.4 & 10.9 &       & VS &                        \\       
  83.643 & $-$67.326 & 277.438 & $-$32.253 & 15 &  2.505 &  37.582 &  5.5 & 14.7 &       & PS &                        \\       
  83.951 & $-$66.115 & 275.998 & $-$32.256 & 21 &  2.575 &  54.075 &  5.7 & 12.4 &  4.68 & VS &  N~63A                 \\
e 84.457 & $-$69.215 & 279.609 & $-$31.737 &216 &  3.405 & 735.446 & 14.3 & 38.3 &  3.02 & VS &  4FGL J0537.8$-$6909, P2 \\  
         &           &         &           &    &        &         &      &      &  3.60 &    &  N~157B                \\
  90.276 & $-$70.583 & 281.013 & $-$29.637 & 38 &  4.885 & 185.637 &  3.4 & 13.2 &  1.24 & -- &  5BZQ J0601$-$7036     \\
\hline
\end{tabular}
}
\label{t:Cl10e6}
\end{table*}

\begin{table*}
\caption{Clusters found in a DBSCAN analysis of the same LMC field. Cluster coordinates, number of events, significance according to Eq.~\ref{eq:signif} and cluster radius are reported. Moreover, association with known sources and/or MST clusters reported in Table~\ref{t:Cl10e6}, with the value of the relative spatial separation.
The two clusters DBS($81.679,-69.000$) and DBS($81.694,-69.235$) are a possible fragmentation of the cluster MST($81.700,-69.129$). }
\centering
\begin{tabular}{rrrrrrrrr}
\hline
RA 	     &	    DEC 	&	 L 	        &	B 			&	N$_\mathrm{evt}$	& S$_\mathrm{cls}$		&	R$_\mathrm{cl}$	& Association	 			& MST $\Delta\theta$ \\
\degr	 &	\degr		&	\degr		&	\degr		&								&				& \arcmin			&			& \arcmin \\
\hline
77.456	 &	$-$64.289	&	274.273	&	$-$35.203	&	19					&  3.58	&	5.52			& 4FGL J0509.9-6417  	& 1.18  \\
80.738	 &	$-$67.932	&	278.334	&	$-$33.264	&	8					&  1.45	&	6.10			& N 44					& 3.53  \\
81.292	 &	$-$69.614	&	280.269	&	$-$32.778	&	14					&  2.86	&	5.39			& N 132D					& 0.32  \\
81.305	 &	$-$65.934	&	275.930	&	$-$33.345	&	7					&  1.73	&	5.00			& N 49B					& 1.85  \\
81.519	 &	$-$66.116	&	276.132	&	$-$33.235	&	6					&  1.72	&	5.06			& N 49					& 1.34  \\
81.679	 &	$-$69.000	&	279.522	&	$-$32.747	&	8					&  1.96	&	5.22			&  $^*$MST($81.700,-69.129$)      & 7.6   \\
81.694	 &	$-$69.235	&	279.797	&	$-$32.704	&	7					&  1.82       &	4.64			&  $^*$MST($81.700,-69.129$)            & 6.4   \\
82.429	 &	$-$68.740	&	279.167	&	$-$32.521	&	10					&  2.05	&	5.93			& 4FGLe					& 2.17  \\
82.702	 &	$-$69.197	&	279.686	&	$-$32.356	&	11					&  2.39	&	5.51			& 4FGLe					& 0.27  \\
83.797	 &	$-$69.479	&	279.951	&	$-$31.934	&	6					&  1.40	&	5.08			& ---	                & 3.02  \\
83.968	 &	$-$66.047	&	275.917	&	$-$32.256	&	12					&  2.25	&	6.10			& N 63A					& 0.43  \\
84.324	 &	$-$69.157	&	279.548	&	$-$31.790	&	76					&  7.13	&	9.98			& N 157B	                & 1.00  \\
85.124	 &	$-$69.346	&	279.730	&	$-$31.487	&	10					&  2.39	&	5.53			& PSR B0540-69	        & 1.67	\\
90.289	 &	$-$70.576	&	281.007	&	$-$29.633	&	19					&  4.02	&	4.20			& 5BZQ J0601$-$7036	 & 0.58	\\
\hline
\end{tabular}
\label{t:dbscan}
\end{table*}

\begin{figure*}  
\includegraphics[width=0.75\textwidth, height=0.50\textwidth]{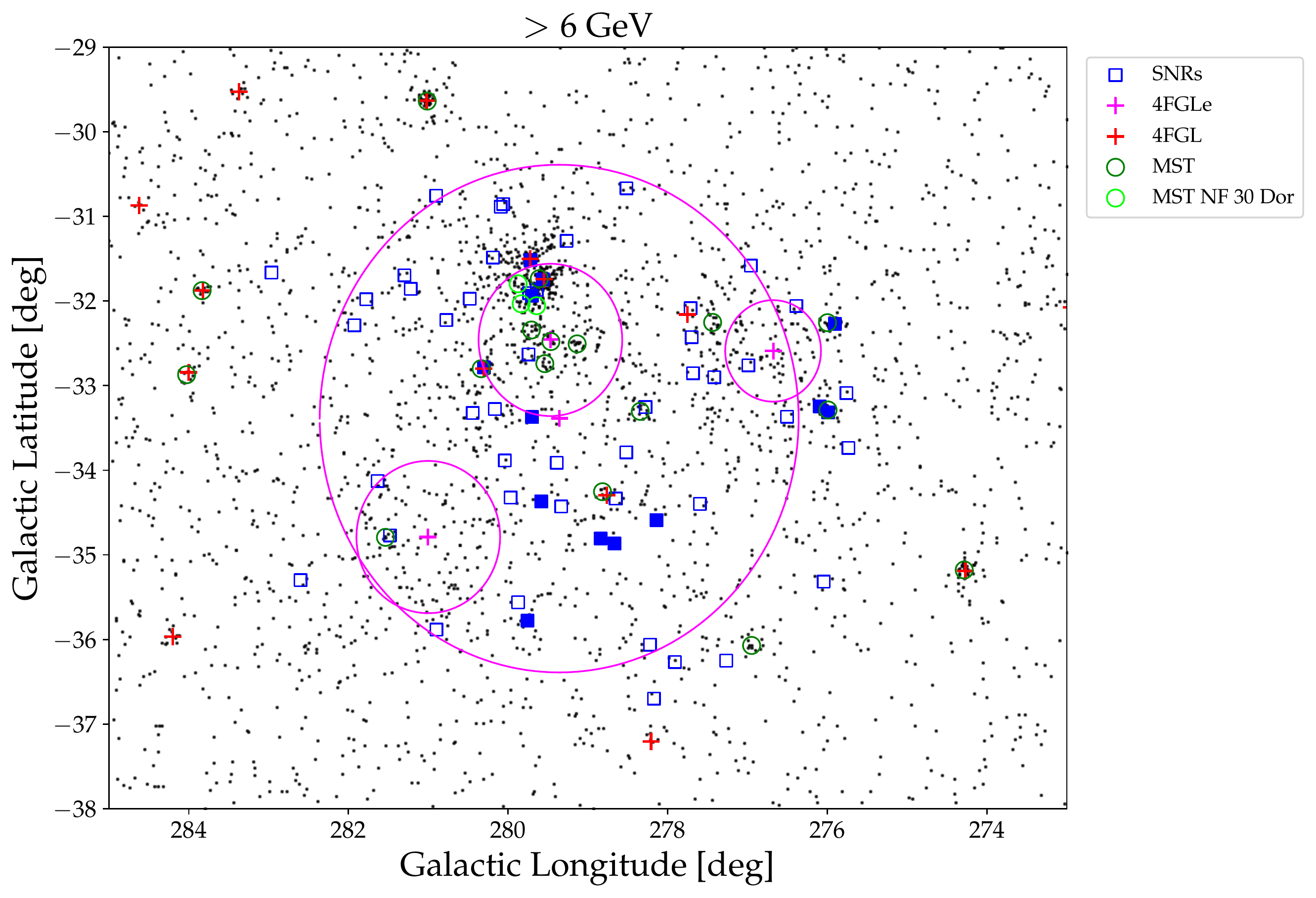}
\includegraphics[width=0.70\textwidth, height=0.50\textwidth]{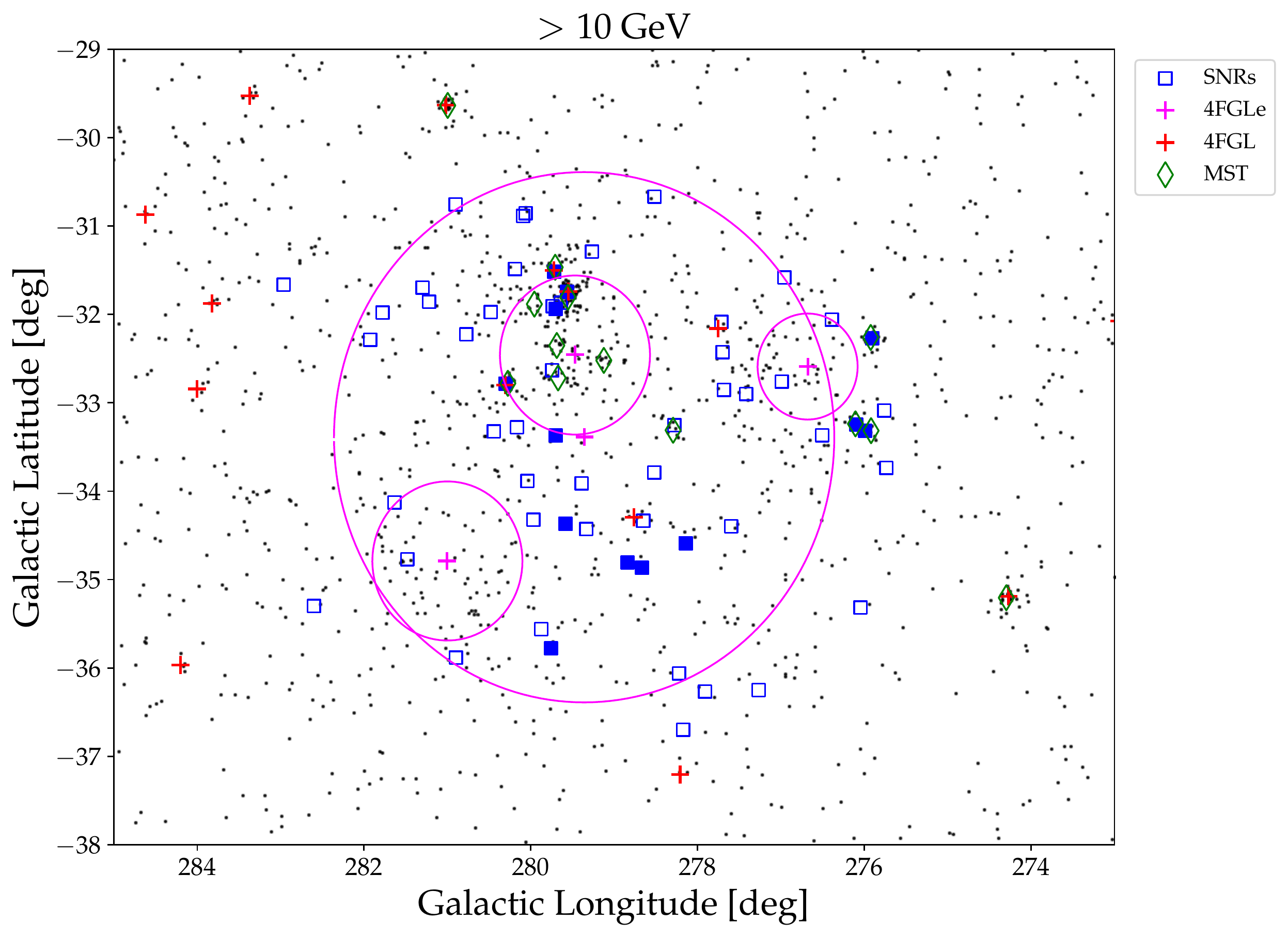}
\caption{
Upper panel: Photon map in Galactic coordinates of the sky region at energies higher
than 6 GeV.
Red and magenta crosses mark the positions of 4FGL-DR2 sources (point-like and 
extended), magenta circles are the sizes of the extended ones; blue squares are the 
SNRs in the \citet{maggi16} catalogue and filled squares are those with an X-ray 
luminosity higher than 10$^{36}$ erg/s;
dark green circles are MST clusters found in the analysis at energy higher
than 6 GeV and $\Lambda_\mathrm{cut} = 0.7$ undetected above 10 GeV; light green 
circles are the clusters found in the region of 30 Doradus complex.
Lower panel: Photon map in Galactic coordinates of the sky region at energies 
higher than 10~GeV.
Source symbols are as in the upper panel; green diamonds are the 
clusters found in the large field analysis at energies higher than 10~GeV and 
$\Lambda_\mathrm{cut} = 0.7$.
}
\label{skymap_f2}
\end{figure*}

\section{Cluster search results}   
\label{s:analysis}

The high energy image of the LMC is extremely complex and the extraction of 
individual sources is not a simple process.
A first analysis of this region was presented by \citet[]{ackermann16} who 
investigated the \fl data obtained in the first six years of observation.
Their maps show a large extended region with a dominant maximum in the region 
of 30 Dor.
These authors detected four sources, labelled from P1 to P4: P1 was undoubtebly 
identified with the pulsar PSR J0540$-$6919 because of his periodic signal, P2 
was found coincident with PSR J0537$-$6910 and the SNR N~157B, P4 with the SNR 
N~132D, while P3 was unassociated and after was identified with the first 
extragalactic $\gamma$-ray binary within the SNR DEM L241 \citep{corbet16}.
\citet{tang18} found a new source in the LMC region at energies lower than 10 
GeV, which was confirmed by MST (see Paper I) search after extending the energy 
range down to 7 GeV.
The most relevant result in Paper I was the discovery of two significant clusters 
associated with the two SNRs N~49B and N~63A, that increased to four the number 
this type of sources in LMC.

On this basis one can reasonably expect that the increase of data would confirm 
these results and provide more evidence for new SNRs.
We therefore applied cluster search to the photon maps at energies above 10 GeV and 6 GeV 
and extracted clusters with a magnitude $M \ge 20$, a value high enough to garantee 
a high confidence level as also recently verified by \citet{campana21}.
The search was then extended to narrow fields for confirming the clusters detected 
in the previous steps and, eventually, for sorting new source candidates.

Centroid coordinates and other interesting parameters of the high significance 
($M \geq 20$) clusters selected at energies higher than 10 GeV applying MST with 
$\Lambda_\mathrm{cut} =$ 0.6 are reported in the first part of Table~\ref{t:Cl10e6}, 
while in the second part lists the clusters found at energies above 6 GeV applying 
$\Lambda_\mathrm{cut} = 0.7$.
In the former part there are 13 clusters, six of 
which have a positional correspondence with sources in the 4FGL-DR2 catalogue 
\citep{abdollahi20,ballet20} and with three of the sources found by \cite{ackermann16}, 
which are indicated by P1, P2 and P4, while P3 was not detected.                                   
All clusters reported in Paper I are confirmed, although the centroid coordinates 
of extended clusters do not match precisely but are within their maximum radii.
In particular, the two SNRs N~49B and N~63A are now found with a safe $M$ value 
and there is a good evidence for high energy emission from some other remnants.
A map of the region with the positions of SNRs, 4FGL-DR2 sources and our clusters 
is given in Figure~\ref{skymap_f2}.
Finally, we note that no cluster was found at a position near to that of the 
unassociated P3 source reported by \cite{ackermann16}: this could be explained 
by its steep spectrum (the reported spectral index is $2.8\pm0.1$) implying a 
low photon number above 6 GeV.

In the latter part of the Table~\ref{t:Cl10e6} are reported 17 clusters found 
in the analysis at energies higher than 6 GeV: essentially all clusters found 
at higher energies are confirmed, with a couple of uncertain associations as 
reported in the individual descriptions (see below).

In the following subsections we describe some of the main properties of individual clusters and their possible counterparts.

The results of the parallel DBSCAN analysis are reported in Table~\ref{t:dbscan}. The detection has been performed selecting photons above 10 GeV, end cutting a circular region with a radius of 5\degr\  around the centre of the same region used for the MST analysis. The value of $K_\mathrm{frac}$ has been set to 3.5, and $\varepsilon$ has been set equal to 0\fdg1, close to the Fermi-LAT PSF at 10 GeV. We notice  an excellent agreement both in the cluster detection and in the photon number implying very close centroid positions.
All the MST clusters have a one-to-one match with those detected by the DBSCAN, with the exception of the cluster MST($81.700,-69.129$). For this source, the MST has detected a cluster with 17 photons, whilst the DBSCAN has detected two clusters ( DBS($81.679,-69.000$) and DBS($81.694,-69.235$) ), with 8 and 7 photons respectively, with a distance of approximately 7\arcmin.

\subsection{Details on MST clusters at energies $>$ 10 GeV}   
\label{ss:clust10G}

\paragraph*{MST(77.440, $-$64.308).}
A high significance cluster with a centroid very close to the present one
was reported in Paper I.
NF analysis confirmed a rich cluster.
It is very likely associated with a 4FGL source and the proposed counterpart is
a radio and X-ray source, likely a background blazar.

\paragraph*{MST(80.630, $-$67.891).}
This cluster is a new detection and its position is close to that of the SNR 
N~44, that is at an angular distance from the centroid less than 4\arcmin, fully 
compatible with the $\gamma$-ray positional accuracy.
This cluster is found in all five NFs where it is included with a value of $M$ 
ranging from 17.5 to 30.4, with the only exception of NF B which however includes 
30 Dor and has a mean photon separation much lower than the other NFs.

\paragraph*{MST(81.284, $-$69.618).}
It is positionally associated with the source P4 of \citet{ackermann16} that 
is also reported in the 4FGL-DR2.
It is only found in NF B and C, with associated clusters having $M$ equal to 
34.5 and 43.3.
It was already reported in Paper I and corresponds to the SNR N~132D.

\paragraph*{MST(81.379, $-$65.928) and MST(81.515, $-$66.093).}
MST analysis found a close pair of clusters: one of them with the centroid nearly
coincident with a previously reported feature in Paper I that was associated with
the SNR N~49B, while the centroid of the other cluster is very close to the near
SNR N~49, at an angular distance of only 6\farcm66, comparable to the typical 
size of clusters.
In Paper I it is reported only a cluster whose association with N~49B was more 
robust with respect to the one with N~49.

The present analysis with the usual separation length gives two nearby clusters: one 
is very close to N~49 with the magnetar SGR 0526$-$66, while the other is closer to 
N~49B at a distance of $\sim$3\farcm7, compatible with the cluster size.
Searches with a higher separation cluster (0.8) give a unique joint cluster 
with a photon number around 16 and $M \approx 40$, 
A cluster appears also in the NF searches with a position intermediate between the 
two remnants.

\begin{figure}  
\includegraphics[width=0.5\textwidth]{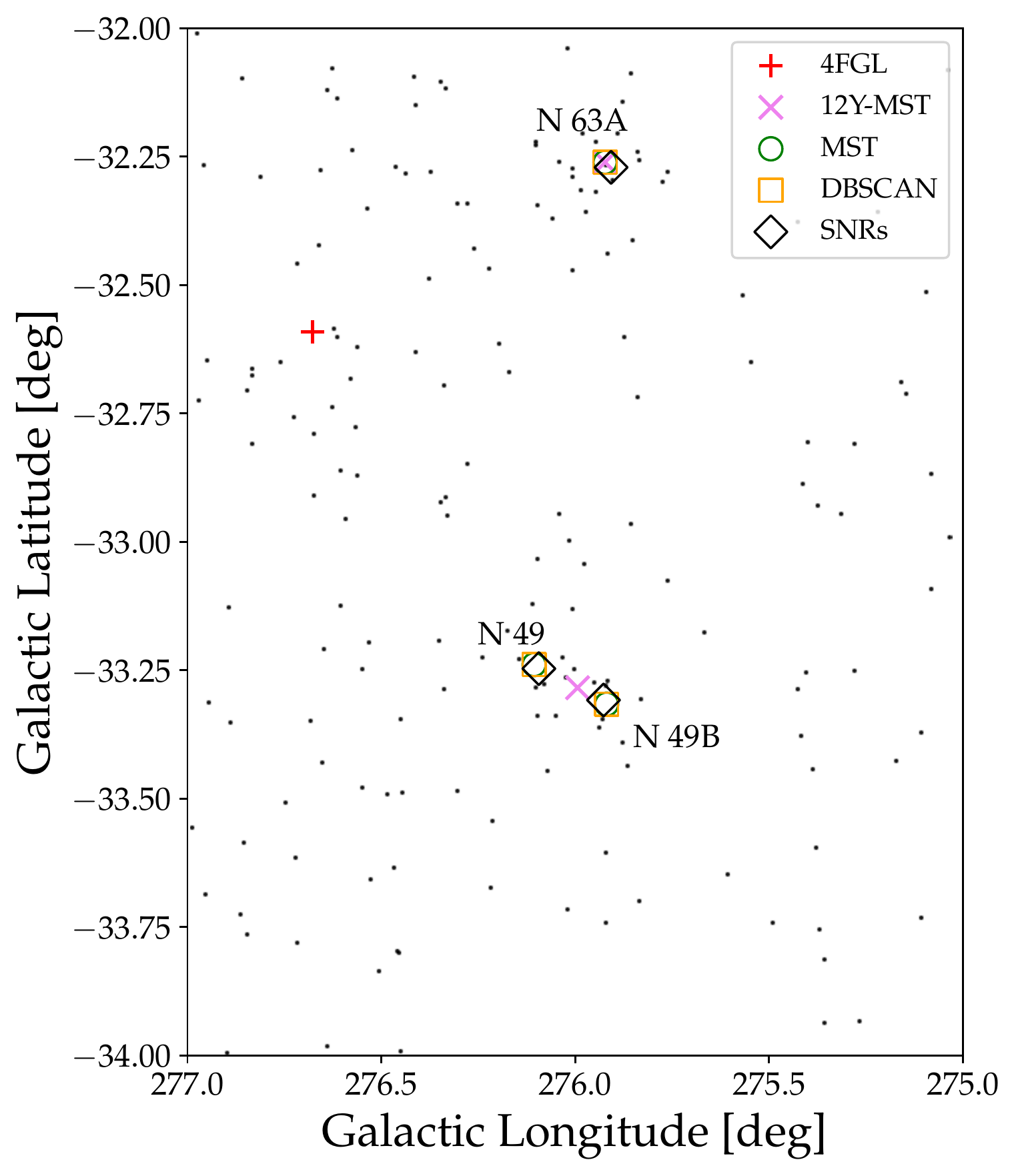}
\caption{
Photon map in Galactic coordinates of the sky region including the SNRs N~63A, N~49 and N~49B at energies higher than 6 GeV.
Open diamonds are the SNRs, large violet crosses are the clusters in the 12Y-MST catalogue, green circles are the MST clusters found in the present analysis as reported in Table~\ref{t:Cl10e6}
above 10 GeV, while orange squares are the DBSCAN clusters reported in in Table~\ref{t:dbscan}. The red cross is only source in the 4FGL-DR2 catalogue detected in this region.
}
\label{f:N49N63}
\end{figure}

A $2\degr \times 2\degr$ size photon map of the region is shown in Figure~\ref{f:N49N63},
where the clusters and SNRs positions are given together with the 4FGL-DR2 sources
and 12Y-MST clusters.
Note that the pair N~49 and N~49B is unresolved in this catalogue, while the source
4FGL J0531.8-6639e is classified as extended and without any reported possible counterpart;
there is an entry in the MST large field analysis for the 12Y-MST catalogue close to 
its position but the parameters are just below the adopted selection thresholds.
It should be emphasised that with the present analysis no definite statement about the separation of these two
features can be given, since the occurrence of only one more photon in the intermediate position might be sufficient to join the two structures in a single cluster.
However, when considering a single connected cluster it 
appears clearly elongated in the direction connecting the two SNRs, suggesting that photons
are likely originated by both objects.

The analysis in NF A above 10 GeV with $\Lambda_\mathrm{cut}$ = 0.6 found only 4
clusters with $M >$ 17, two of which corresponding to the ones in the complete field, 
while in the energy band higher that 6 GeV only an unresolved feature with 
$M = 28.4$ is found, likely due to the broader PSF.

\paragraph*{MST(81.700, $-$69.129).}
A cluster near this position is reported in Paper I only for 
$\Lambda_\mathrm{cut} = 0.5$,
but in the present search it is found at the two cut distances employed, with 
different photon numbers and $R_\mathrm{max}$ values.
It is located near the centre of the extended source 4FGL-DR2 J0530.0$-$6900e 
which contains 30 Dor and N~157B, and therefore with a high local photon density.
It is close to SNR B0528$-$692, but the separation is higher than 6\arcmin, 
however the median radius is over this distance; the analysis with 
$\Lambda_\mathrm{cut}$ = 0.5 gives 
a cluster of only 6 photons and the distance to the remnants is slightly lower 
than $R_\mathrm{max}$.
NF B and C searches do not confirm a stable cluster and give a marginally 
significant feature in the extended 4FGL-DR2 source.
This association shall be considered uncertain.

\paragraph*{MST(82.443, $-$68.704) and MST(82.714, $-$69.199).}
These two nearby rather compact clusters are located at about 11\arcmin\ and 
18\arcmin\ from the centre of the same 4FGL-DR2 extended source as the 
previous one.
No known SNR is close enough to their centroids to be associated with, however, 
there is a long list of stars and other sources within a distance of 3\arcmin, 
and the selection of a reliable counterpart candidate is a very difficult task.
NF analysis is in agreement with this result and therefore one cannot exclude 
that the extended feature is due to some close unresolved structures.

\paragraph*{MST(83.939, $-$69.490).}
This is a quite peculiar cluster because it is found with 
$\Lambda_\mathrm{cut} = 0.6$, but not with a lower value.
The clustering factor is rather low, but not enough to produce a magnitude 
below the threshold.
For $\Lambda_\mathrm{cut} = 0.7$ a cluster of 20 photons and a very close 
position is sorted but with a lower $g$ and $R_\mathrm{max} =$ 23\farcm5; 
for $\Lambda_\mathrm{cut} = 0.8$ it is merged in a single and very large 
feature with $R_\mathrm{max} \approx$ 1\degr and is connected to the large 
30 Dor nebula.
We did not find a good candidate for a possible counterpart.
Note that the cluster is a region containing some SNRs: at an angular 
separation of 13\farcm3 there is the very young remnant of SN 1987a, the 
bright source N~157B is at 21\farcm7, and the two other remnants B0540$-$693 
and B0534$-$699 are at 25\farcm2 and 27\farcm2, respectively.
The MST search applied to NF B gave three close clusters of which only one
has a remarkably high $M$.
We therefore cannot exclude that the cluster is originated by a photon 
density fluctuation in a region dominated by a high diffuse emission or 
if it is produced by the superposition of the peripheral radiation of 
some sources.

\paragraph*{MST(83.955, $-$66.051).}
This cluster is one of the most stable features in our NF analysis: a 
cluster at the same position is, in fact, found for different choices 
of the selection parameters, energy ranges and searching regions.
It is clearly associated with SNR N~63A as was already discussed in Paper I.
The photon map showing the cluster associated with this object is
also shown in Figure~\ref{f:N49N63}.

\paragraph*{MST(84.278, $-$69.160).}
As already found in Paper I, this cluster is the richest and the densest 
one in the sample. 
It has a correspondent source in the 4FGL-DR2 (P2 source), safely associated 
with the SNR N~157B:
the angular separation between the cluster centroid and the remnant is 
3\farcm5, lower than in Paper I and well compatible with instrumental PSF 
at these energies. 
However, the large value of maximum radius, close to 25\arcmin, much higher 
than the X-ray size of the PWN, which according to \citet{wang01} is around 
0\farcm5, suggests the presence of an extended component. 
The analysis performed applying $\Lambda_\mathrm{cut} = 0.5$ gives 
an even denser cluster with only a slightly lower photon number and 
$R_\mathrm{max} = $19\farcm4. 
These results confirm the this feature is embedded in an extended emission, 
as already reported in Paper I.

\paragraph*{MST(85.186, $-$69.329).}
A similar feature was already reported in Paper I.
It corresponds to another one of the four sources detected by 
\citet[]{ackermann16} and its counterpart is SNR B0540$-$693 with the young 
pulsar PSR J0540$-$6919.
This source is located in one of the brightest regions and for this reason it
is not included in the list of clusters detected above 6 GeV, since it is 
unresolved from the near extended cluster which has a radius of about 
38\arcmin.
It is detected in NF B, but with a variable $M$, depending on the adopted 
$\Lambda_\mathrm{cut}$: with a separation length equal to 0.6 the corresponding 
cluster has 20 photons and $M = 59.217$, while for $\Lambda_\mathrm{cut} = 0.7$ 
these values are reduced to 8 photons and $M = 18.035$
For this reason one cannot consider this a VS cluster, but its very high 
magnitudes in some searches correspond to a very significant feature.

\paragraph*{MST(90.281, $-$70.568).}
This rich cluster corresponds to a source in the 4FGL-DR2 catalogue.
It is outside of all considered NFs.
It was already found in Paper I as a high significance cluster, clearly 
associated with the background flat spectrum radio quasar 5BZQ J0601$-$7036 
(PKS 0601$-$70).

\subsection{Details on individual clusters at energies $>$ 6 GeV}   
\label{ss:clust6G}

In the following we report the main properties of only the clusters not found 
above 10 GeV.

\paragraph*{MST(74.306, $-$66.225).}
It is the cluster with the highest $M$ in this sample; it is found above 8 GeV
but not above 10 GeV, that could imply a rather soft spectrum with a cut-off
around this energy.
Its position is rather peripheral and does not correspond to any known SNR.
The search for a possible counterpart indicates at a rather short angular 
separation a SUMSS radio source with an optical $B$ magnitude around 20.0, 
corresponding also to a ROSAT X-ray source.
It might be associated with a background blazar, but further observations 
are necessary to confirm its nature.

\paragraph*{MST(74.875, $-$70.169).}
This cluster with a high $M$ value above 6 GeV is located at a rather short 
angular separation from the SNR N~186D, well compatible with the angular 
resolution at these energies.
In the large field analysis there is a cluster with only 4 photons and 
$M \approx 15$.
It is located in the same area covered by the LMC-FarWest.
The association with the SNR N~186D appears likely although not firmly 
established.

\paragraph*{MST(77.934, $-$68.137).}
This cluster is in the large field analysis with a value of $M$ just below the 
adopted threshold. 
It has a clear correspondence with a 4FGL-DR2 source, associated in this catalogue 
with a background AGN of uncertain classification.
There is a faint SNR in the \citet{maggi16}  list at the angular distance of 
5\farcm8, slightly higher than $R_\mathrm{max}$.
We can consider this detection as safely established, while 
its nature is not well understood.

\paragraph*{MST(78.936, $-$72.681).}
There is a very well positional correspondence with a 4FGL-DR2 source whose 
candidate counterpart is a flat spectrum radio source (PKS 0517$-$726) classified 
as a blazar of uncertain type (see also \citealt{dabrusco19}).
A $\gamma$-ray flare with a rather hard spectrum was reported in December 2018 
\citep{rani18}.
Considering that it is undetected above 10 GeV, while a low $M$ cluster is 
found at energies higher than 8 GeV, one can expect the occurrence of a sharp 
cutoff around this energy.

\paragraph*{MST(82.317, $-$72.740).}
Its coordinates are very close to those of a 4FGL-DR2 associated with the flat 
spectrum radio source (PKS 0530$-$727) classified as a blazar of a mixed type
between BL Lacs and flat spectrum radio quasars \citep{dabrusco19}.
A cluster with $M = 17.803$ is found at energies higher than 8 GeV, while no 
feature is found in the MST search above 10 GeV.

\paragraph*{MST(83.643, $-$67.326).}
This is a rather rich cluster without any corresponding feature at energies 
higher than 8 GeV.
In the NF analysis it is divided into a couple of near clusters which are 
joined together when a slightly higher $\Lambda_\mathrm{cut} = 0.8$ is chosen; 
again, no feature is found at higher energies.
The cluster is not very stable and its detection is critically dependent on 
the selection parameters. 
Likely it could be originated by local density fluctuations in a relatively 
bright region.

\subsection{A new cluster from NF analyses}                   
\label{ssct:Enrancl}

Complementary cluster NF searches to verify and to extend the results 
above 10 GeV, provided only a new cluster having a magnitude higher 
than 20, indicated as MST(83.553, $-$69.203) using the same nomenclature 
as before.
At energies higher than 6 GeV, this cluster has only 6 photons but a quite high 
$g = 3.944$ that implies a magnitude $M = 23.655$.
At higher energies no significant cluster is found.
It is well embedded in the 30 Dor complex and consequently its significance 
cannot be safely established, because it is not easy to exclude a local 
background fluctuation.
A tentative analysis down to 4 GeV gave clusters with a well compatible 
position but a magnitude critically dependent on $\Lambda_\mathrm{cut}$.
Moreover, its angular distance from SN 1987A is about 23\arcmin, large 
enough for a possible association.

In any case we searched for some possible interesting candidate counterpart 
and found a radio (PMN J0536$-$6909) and X-ray source (2XMM J053438.1$-$691003)
at an angular separation lower than 3\arcmin, whose nature is not known.


\section{Bright SNRs in LMC }  \label{s:snr}
 
In Table \ref{t:Cl10e6} some of the reported detections correspond to SNRs with 
high X-ray luminosity.
We then extracted from the \citet{maggi16} list a subsample of 13 bright sources 
having a luminosity higher than 10$^{36}$ erg s$^{-1}$.
Six remnants of this subsample were found associated with significant clusters, as discussed in the previous sections.
In the same table we reported also the flux densities at 1 GHz from the \citet[]{bozzetto17} catalogue: 12 sources (20\%) of 59 confirmed SNRs are above this value, and only 5 (about 10\%) have a flux density higher than 1 Jy. 
All these radio bright remnants are detected in our analysis and only N~49B has a flux density lower than 1.0~Jy. 
In the third section of this Table we reported the
other four SNRs belonging to the subset of radio flux density higher than 0.5 Jy, which have a much older age estimates and rather low X-ray luminosity.
No correlation is found between the radio and $\gamma$-ray fluxes (the linear correlation coefficient is 0.07), however, these results suggest that the total energy output of the detected SNRs is remarkably higher than that of other sources.
We then investigated in more detail if some of the remaining remnants, 
reported in Table \ref{t:brightsnr}, may be found among the low significance 
clusters, in energy ranges higher than 6 GeV.
In particular we applied MST to small fields of $2\degr\times2\degr$ 
approximately centred on the various remnants and no interesting cluster was 
found near the coordinates of these remnants.
We can conclude that, if these sources are gamma-ray loud, their spectrum should
have a cut-off at energies lower than 6 GeV.

\subsection{Evaluation of $\gamma$-ray fluxes}

Given the very complex background structure in the LMC, to extract the 
$\gamma$-ray fluxes from the selected source candidates using the standard ML 
is rather difficult, since this method is extremely sensitive to the spectral 
model and adopted background structure.

We opted, therefore, to perform a simplified estimate of the $\gamma$-ray flux  
using the aperture photometry 
method\footnote{\url{https://fermi.gsfc.nasa.gov/ssc/data/analysis/scitools/aperture_photometry.html}},
in which the flux is computed by essentially dividing the photon number 
in a region of interest by the exposure in same region.
For each source in Table~\ref{t:brightsnr} which corresponds to a MST cluster, 
the  $\gamma$-ray flux has been estimated in the 3--300 GeV band, using as a 
source region of interest a circle of radius $0.9R_\mathrm{max}$. 
The background has been estimated in a circular region of radius 
$1.5R_\mathrm{max}$ likewise centred on the MST coordinates. 
Net source fluxes are therefore computed by rescaling and subtracting the 
expected background. 
It should be emphasised that a cross-contamination between source and 
background cannot be excluded, but the resulting estimate should be a reliable
indication of the actual  $\gamma$-ray flux.
The last column in Table~\ref{t:brightsnr} reports the evaluated fluxes. 
This simple method cannot be safely applied at energies below a few GeV because the radius of the PSF is larger and the evaluation of the local background is much more uncertain. A preliminary study of the diffuse emission in LMC is necessary. For the same reason, reliable estimates of $\gamma$-ray spectral indices cannot be performed.
The comparison between the aperture photometric values and those obtained by
ML analysis for N~157B \citep{ackermann16} and N~63A \citep{campana18} are in a 
satisfactory agreement within the reported standard deviations, with only a possible 
underestimate not higher than 25\%.
Thus we conclude that for the previously detected SNRs the fluxes obtained here are 
consistent with those derived from the standard ML analysis.

\begin{table*}      
\caption{SNRs in the LMC region with a X-ray luminosity in the band 0.3--8 keV higher 
than 10$^{36}$ erg~s$^{-1}$ from the \citet{maggi16} list. Radio flux densities at 1 GHz are from \citet{bozzetto17}.
}
\centering
{\small
\begin{tabular}{lclrrlrlrl}
\hline
 SNR  & MCSNR &  X size   & $L_X$~~~~~        & $S_\mathrm{1\,GHz}$ & Type & Age & MST/DBSCAN & $\gamma$-ray flux$^*$ & Interaction w/  \\
      &          &  $'$~~~  &  10$^{35}$ erg~s$^{-1}$ & Jy~~~ &     &  kyr & detection & 10$^{-11}$ ph cm$^{-2}$~s$^{-1}$   & dense CSM \\
\hline
   &  \\
\hline
N~132D        &  J0525$-$6938 & 2.1  & 315.04 & 5.26 & CC    & 2.4      &  yes & 9.2 $\pm$ 2.3   & yes  \\ 
N~63A         &  J0535$-$6602 & 1.4  & 126.00 & 1.86 & CC    & 3.5      &  yes & 8.5 $\pm$ 1.9   & possible  \\ 
B0540$-$693   &  J0540$-$6920 & 1.2  &  87.35 & 1.03 & CCP   & 1.14     &  yes & 20.9 $\pm$ 4.5  & no \\ 
N~49          &  J0526$-$6605 & 1.4  &  64.37 & 1.66 & CC    & 4.8      &  yes & 5.7 $\pm$ 2.6$^\dagger$ & yes \\ 
N~49B         &  J0526$-$6605 & 2.8  &  38.03 & 0.63 & CC    & $\sim 10$&  yes & 6.8 $\pm$ 3.2$^\dagger$ & possible\\
N~157B        &  J0537$-$6910 & 2.0  &  15.00 & 2.88 & CCP   & $\sim 5$ &  yes & 30.9 $\pm$ 4.5  & yes \\
SNR1987A      &  J0535$-$6916 & 0.03 &  27.39 & 0.82 & CC    & 0.03     &  confused &            & yes \\
\hline
B0453-685     &  J0453$-$6829 & 2.0  &  13.85 & 0.21 & CC    & 12.0     &  no  &               & no \\
N~23          &  J0505$-$6802 & 1.6  &  26.25 & 0.39 & CC(?) & 4.6      &  no  &               & no \\ 
N~103B        &  J0509$-$6843 & 0.5  &  51.70 & 0.58 & Ia    & 0.86     &  no  &               & no \\ 
DEM L71       &  J0505$-$6753 & 1.3  &  44.59 & 0.01 & Ia    & 4.4      &  no  &               & no \\
B0519$-$690   &  J0519$-$6902 & 0.6  &  34.94 & 0.13 & Ia    & 0.6      &  no  &               & no \\
B0509$-$67.5  &  J0509$-$6731 & 0.53 &  16.51 & 0.10 & Ia    & 0.4      &  no  &               & no \\
\hline
B0450-709     &  J0450$-$7050 & 5.67 &  0.59  & 0.69 & CC(?) &	$\sim 70$&  no  &               & no \\	
DEM L316B     &  J0547$-$6942 & 3.17 &  1.47  & 0.73 & CC(?) &	40.5     &  no  &               & yes \\	
DEM L316A     &  J0547$-$6941 & 3.17 &  1.26  & 0.52 & CC(?) &	33.0     &  no  &               & yes \\	
DEM L328      &  J0550$-$6823 & 5.20 &  1.22  & 0.65 & CC(?) &	         &  no  &               & possible \\	
\hline
\end{tabular}
}

($^*$) Photon flux evaluated in the 3--300 GeV energy band. --  ($^\dagger$) Values possibly 
affected by the nearby SNR. -- CC = Core collapse; CCP = Core collapse Plerion
\label{t:brightsnr}
\end{table*}

\section{Discussion and conclusion} \label{s:conclusions}

The MST and DBSCAN methods applied to the complex region of LMC provides a new evidence
for high energy emission from some SNRs.
It is well known that this small close galaxy is rich of SNRs concentrated
in a sky region of few square degrees.
There is also a strong diffuse emission likely originated by a local component of
cosmic rays interacting with the interstellar gas of HII regions, in particular
in the 30~Dor nebula.
This diffuse component can present some local emission fluctuations which 
make difficult the identification of sources against this non uniform
background.

We considered in our analysis photons in the two energy ranges defined by 
the two lower limits of 6~GeV and 10~GeV, in which the Galactic diffuse 
emission is low enough to allow a safe extraction of clusters.
The analysis was also applied to small regions to verify the stability of selected clusters. Some of these regions exclude 30~Dor 
to avoid the possibility of a reduced capability to extract genuine features given the high photon density around this source.

The capability of the MST method for extracting clusters in $\gamma$-ray 
images is described in the paper on the recent catalogue of candidate sources 
at Galactic latitudes higher than 20\degr\ \citep{campana21}.
The validity of this approach is confirmed by the present analysis, in which we found high 
significance clusters corresponding to the sources detected by 
\citet{ackermann16} and also those reported in the 4FGL-DR2 catalogue, 
with the exception of those with a soft spectrum and consequently a quite low high energy flux.

DBSCAN was already successfully applied to $\gamma$-ray cluster detection \citep{2013A&A...549A.138T}, and the estimated significance is highly correlated with $TS$ provided by Fermi-LAT maximum likelihood analysis. The excellent agreement with the results provided by the MST analysis supplies a further validation for the robustness of the detected clusters. The only two clusters detected above 10 GeV, and not present in the MST analysis, result likely from the fragmentation of the source  MST($81.700,-69.129$), as confirmed by the close position and total number of detected photons.

From the results given in Table~\ref{t:Cl10e6} it is possible to extract a final sample of 20 clusters, of which 11 are present in the two energy ranges:
7 clusters are found above 6 GeV, but not above 10 GeV, and only 2 above 10 GeV.
Considering the possible counterparts, 5 clusters might be associated with background
blazars, but only one of them is safely confirmed as FSRQ while the nature 
of the remaining 4 is uncertain; 7 clusters are clearly associated with SNRs, 
one of which harbouring a pulsar.
In Table~\ref{t:brightsnr} we listed the 13 brightest SNRs in the explored region, with an X-ray 
luminosity higher than $5 \times 10^{36}$ erg s$^{-1}$. Six of them are 
associated with clusters, and particularly the 4 with the highest luminosity.
The only cluster in this bright sample not associated with a SNR is 
MST(80.630, $-$67.891) whose candidate counterpart is N~44.
This remnant has an X-ray luminosity of $9\times10^{34}$ erg~s$^{-1}$ and is 
well separated from the near bright 30~Dor region.
It is interesting that the cluster at energies higher than 6~GeV has about 
twice the photons found above 10~GeV, a result confirmed applying a 
separation length of 0.6$\Lambda_m$ and also by the NF searches.
This suggest that it has a rather soft spectrum.
 
\begin{figure*}  
\includegraphics[width=0.8\textwidth]{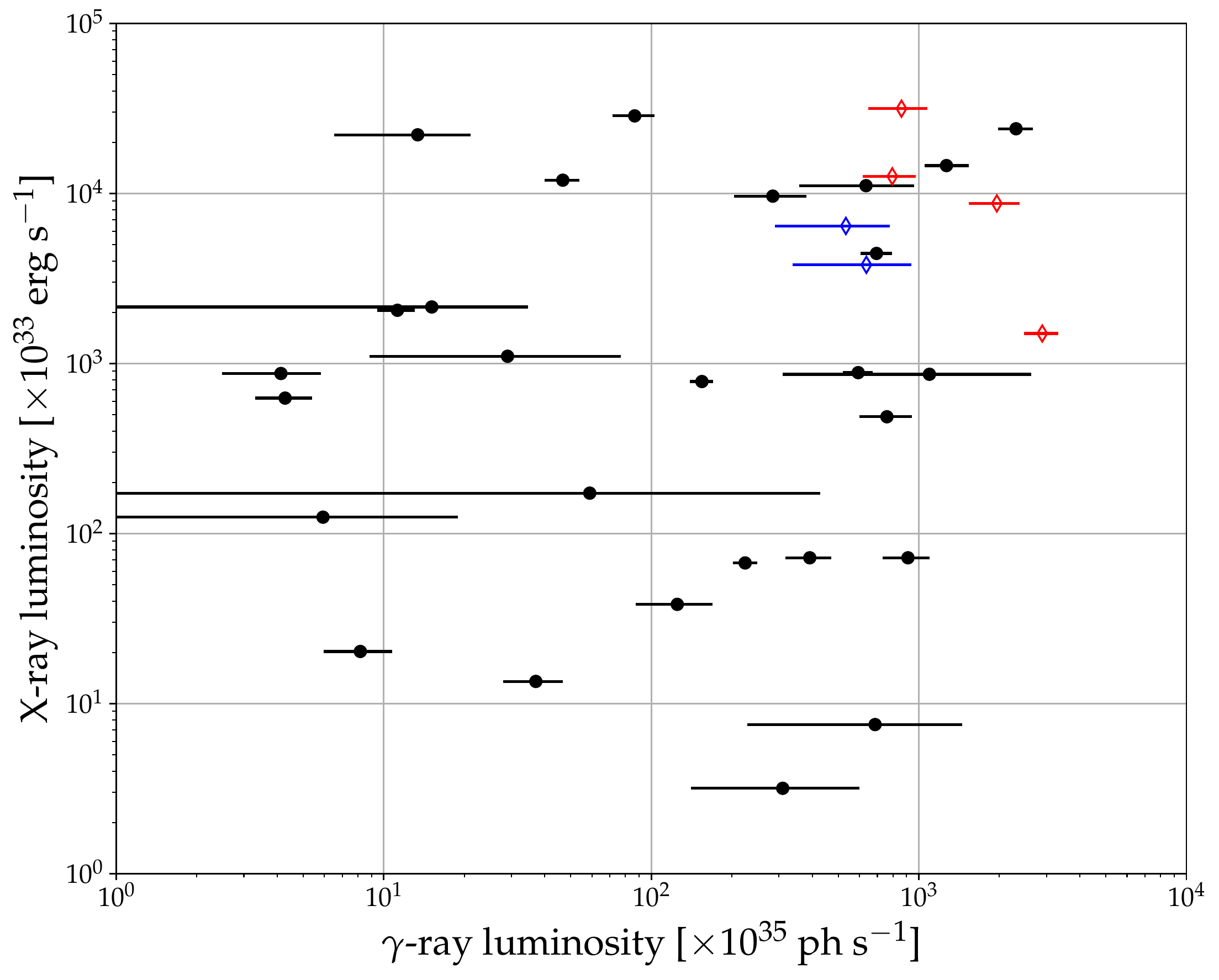}
\caption{Unabsorbed X-ray luminosity of SNRs (0.3--10~keV band) versus the corresponding  $\gamma$-ray luminosity, as measured by \emph{Fermi}-LAT in the 1--100~GeV band. Luminosities for N~132D, N~63A, B0540-693, N~157B (red diamonds), and for N~49 and N~49B (blue diamonds) were derived from the values reported in Table \ref{t:brightsnr} assuming a photon index $\Gamma=2$ and a distance of 51~kpc. See Appendix A for details.}
\label{f:X-GammaSNR}
\end{figure*}

 Figure \ref{f:X-GammaSNR} shows the X-ray luminosities (in the 0.3--10~keV band) of the SNRs detected with \fl as a function of their $\gamma$-ray luminosity in the 1--100~GeV band. Galactic SNRs are indicated by black crosses, the LMC SNRs N~132D, N~63A, B0540-693, N~157B by red crosses, while blue crosses mark N~49 and N~49B (see Appendix~\ref{app:XvsG} for details). 
 $\gamma$-ray fluxes were extrapolated at 1 GeV assuming a photon index $\Gamma=2$, as in the usual prescription of the \fl collaboration for data analysis; this value
 is slightly harder than the mean value of the Galactic SNRs, which according to \citet[]{aaa16} is equal to 2.5.
 The ranges of X-ray and $\gamma$-ray luminosity derived for LMC SNRs are nicely consistent with that of Galactic SNRs: in fact, about one third of Galactic remnants emit in the X-ray or $\gamma$-ray band with a power comparable to those detected in LMC, though we notice that only five sources are among the brightest remnants in both energy bands. Our new detections indicate that the subsample of SNRs in LMC has energetic properties similar to the Galactic ones, even considering the bias toward high fluxes associated with the distance of the LMC and its large $\gamma$-ray background.

\subsection{Properties of some detected SNRs}                   
\label{ssct:snrprop}

In the following we describe the main properties of confirmed or newly detected SNRs in our analysis.

\paragraph*{N~132D.}
This is the remnant of a core-collapse SN with an age since explosion of $\approx$2400~years \citep{2020ApJ...894...73L}. It is the brightest X-ray emitting SNR in LMC \citep{1982ApJ...255..440C, 1993ApJ...414..219H, 1997A&A...324L..45F, 2007ApJ...671L..45B} and its emission in the high energy $\gamma$-ray and TeV ranges is firmly established \citep{hess21,ackermann16}. According to the model considerations developed by \cite{hess21}, the $\gamma$-ray emission exhibits a dominant hadronic component formed by a diffusive shock acceleration. It is possible that the remnant is in interaction with a complex of molecular clouds (e.g., \citealt{2018ApJS..237...10D, 2020ApJ...894...73L}) partially within the spatial confidence region of the source. The photon flux estimated by means of aperture photometry (see Table \ref{t:brightsnr}) is substantially in a good agreement with the one derived from the power-law spectrum with a photon index $\Gamma = 1.4$ given by \citet{ackermann16} that is $1.2\times10^{-10}$ ph~cm$^{-2}$~s$^{-1}$ in the same energy range.

\paragraph*{N~63A.}
This is a bright core-collapse SNR embedded within the large HII region N63, associated with dense molecular clouds, shock-ionised gas, and photoionised gas. Its age is estimated to be about 3500 years (\citealt{sano19}) or within the range $2000-5000$~yr (e.g., \citealt{1998ApJ...505..732H, warren03}). This source is the second brightest SNR in LMC and was observed from radio to X-rays \citep{warren03, williams06, caulet12, sano19}.
Its estimated X-ray luminosity is $4\times 10^{34}$ ergs s$^{-1}$, probably due to an embedded PWN (\citealt{warren03}). 
The first detection of a $\gamma$-ray emission from this bright SNR was already reported in Paper I and was validated by the ML analysis, with $\sqrt{TS} = 9$
and a flux above 1 GeV equal to $(3.4 \pm 0.6) \times 10^{-10}$ ph~cm$^{-2}$~s$^{-1}$.
It is located in near the border of the central high density region of LMC (see
Figure \ref{skymap_f2}) and therefore it is not highly affected by background 
fluctuations. \citet{sano19} assumed that the $\gamma$-ray emission is dominated by hadronic
component and estimated the total energy output in cosmic-ray protons and the
magnetic field which resulted roughly consistent with the Galactic SNRs.

\paragraph*{B0540$-$693.}
This is a composite high luminosity remnant of a core-collapse SN that exhibits a segmented shell emission with an inner bright PWN (PSR J0540$-$6919). Its estimated age is 1140 years (\citealt{2008ApJ...687.1054W}). 
The X-ray emission structure originates from two ``arcs'' near the E and W boundaries and the spectral distribution implies a synchrotron origin \citep{park10,mcentaffer12} from electrons accelerated in the supernova shock. There is no evidence of interaction of the remnant with dense clouds of the ISM. We argue that the high level of $\gamma$-ray emission detected may originate from the PWN.

\paragraph*{N~49.}
Another bright remnant of a core-collapse SN, among the brightest SNRs in LMC. Its estimated age is 4800 years (e.g., \citealt{2012ApJ...748..117P}). A magnetar (SGR B0526$-$66) is clearly detectable as a bright point-like X-ray source \citep{park12,zhou19} closer to the North boundary than to the centre. The magnetar has been recently detected by NuSTAR \citep{park20} in a quiescent state up to 40 keV; above about 10 keV the emission is dominated by a power-law component.
The physical association of the magnetar and the SNR, however, is not yet safely 
established, and the possibility of a casual spatial coincidence cannot be ruled out.
We do not know if this magnetar exhibits or not a quiescent emission in the high
energy $\gamma$-rays, but no detection in the TeV range has been reported up to now \citep{komin19}.
Multi-wavelengths observations show clear evidence of interaction of the remnant with dense clumpy interstellar clouds on the eastern side of the remnant (\citealt{1992ApJ...394..158V, 1997ApJ...480..607B}). The interaction leads to bright emission in radio, optical, infrared, ultraviolet, and X-ray bands (e.g., \citealt{2003ApJ...586..210P, 2004AJ....128.1615S, 2006AJ....132.1877W, 2007ApJ...655..885R, 2007AJ....134.2308B, 2010A&A...518L.139O, 2012ApJ...748..117P}).

\paragraph*{N~49B.}
This remnant is a core-collapse SNR with an estimated dynamical age of $\approx 10000$ years (\citealt{1998ApJ...505..732H}). Its X-ray structure was investigated by \citet{park17} who supported the possibility of a large mass progenitor.
The X-ray luminosity of this SNR, however, is about the 60\% of that of N~49; in any case it is certainly among the brightest remnants in the LMC. The angular separation between N~49 and N~49B is comparable with the $\gamma$-ray resolution at energies of
$\approx$ 10 GeV or higher but they cannot be resolved at lower energies.
Nevertheless, the elongation of the cluster in the direction connecting the two SNRs
is an indication that both are responsible for the detected feature.
\cite{1988AJ.....96.1874C} evidenced a close proximity of N~49B with the HII region DEM 181 and a possible association with a molecular cloud (see, also, \citealt{1988ApJ...331L..95C}). In particular, the superposition of faint and bright features suggests that the remnant is expanding through a cloudy ISM. In light of this, \cite{1988AJ.....96.1874C} suggested a Population I progenitor for N49B.

\paragraph*{N~157B.}
This is the remnant of a core-collapse SN, characterized by a large physical size in both radio and X-ray bands (e.g., \citealt{1998ApJ...494..623W, 2000ApJ...540..808L}). Its age is not well constrained; however, according to the average ionization age inferred from the analysis of X-ray observations, the estimated age of the remnant could be $\approx 5000$~years (\citealt{2006ApJ...651..237C}). The remnant contains the young pulsar PSR J0537$-$6910, an ultrafast X-ray pulsar with a period of 16 ms, surrounded by a X-ray–bright nonthermal PWN (\citealt{1998ApJ...499L.179M}). The remnant appears to expand through the hot low-density medium of a surrounding superbubbles formed by the young OB association LH 99 (\citealt{1970AJ.....75..171L, 1992AJ....103.1545C}); this may explain the unusual large size of the remnant. On the other hand, the comparison between optical and X-ray observations revealed that the explosion site is located close to a dense cloud which led to a reflection shock that is now interacting with the PWN (\citealt{2006ApJ...651..237C}).

\paragraph*{N~44.} This is the only candidate counterpart of clusters that is not directly associated with a SNR. In fact, N~44 is a rather complex nebula identified as an active star forming region with superbubble structure, comprising a large complex of H II regions (\citealt{1986ApJ...306..130K, 2012ApJ...755...40P, 2021AJ....161..257K}). The nebula is populated by three OB associations (LH47, 48 and 49; \citealt{1970AJ.....75..171L}) and several massive O-type stars (see \citealt{2021AJ....161..257K} and references therein). Interestingly, N~44 also hosts SNR 0523-679 (\citealt{1993ApJ...414..213C}), a remnant showing a high $\alpha$/Fe ratio as expected from core-collapse SNe and with a total X-ray luminosity of $5.73 \times 10^{35}$~ergs~s$^{-1}$ (\citealt{2011ApJ...729...28J}). Given the complexity of the region and the wealth of high-energy objects in N~44, we cannot identify the object mainly responsible for the detected cluster.

\bigskip

We note that the SNRs associated with our detected clusters are all remnants of core-collapse SNe and the vast majority of them show clear evidence of interaction with a dense ambient medium. The only exception is B0540$-$693 which, however, hosts the young and energetic pulsar PSR J0540$-$6919 which may be responsible for the intense $\gamma$-ray emission. All other SNRs in Table~\ref{t:brightsnr} cannot be uniquely associated with any of our clusters. Of these, the very young remnant of the core-collapse SN 1987A is the only one for which a strong interaction with a highly structured CSM (consisting of an HII region and a dense ring) is known. This remnant, however, was not considered as a possible counterpart of our clusters, as it is located in a very complex and rich area, so a unique association is not possible. Furthermore, \cite{2018ApJ...867....7S} suggested that the SN originating N~103B exploded in a bubble driven by the accretion wind from the progenitor system. Thus, N~103B is now interacting with the dense gas wall and the observations show  evidence of strong interaction between the SN ejecta and the dense CSM (\citealt{2014ApJ...785L..27Y}). Indeed all other remnants show no evidence of interaction with a dense CSM. We also note that most of these remnants originate from Type Ia SNe (N~103B, DEM~L71, B0519$-$690, B0509$-$67.5), with the exception of N~23 (for which, however, there is no consensus for its type of explosion, a core-collapse or Ia; \citealt{2006ApJ...653..280H}) and B0453$-$685 (a fairly old SNR with an age of $\approx$12000~years, which propagates through a low-density medium and shows an almost circular shape with no hints of interaction with clouds; \citealt{2012A&A...543A.154H}). 
More detailed observational analyses would be useful for improving the knowledge of the high energy emission from these objects, in particular for comparing their spectral distribution in different frequency bands. However, in the light of available information, we can conclude that the counterparts of our clusters are only core-collapse SNRs mostly interacting with a dense medium. This evidence may indicate a hadronic origin for their $\gamma$-ray emission.

\section*{Acknowledgments}
We are grateful to the referee D. Uro\v sevi\'c for his useful comments.
We acknowledge use of archival Fermi data. 
F.B., M.M. and S.O. acknowledge financial contribution from the PRIN INAF 2019 grant ``From massive stars to supernovae and supernova remnants: driving mass, energy and cosmic rays in our Galaxy''.

\section*{Data availability}
Data used in this paper are available on request.



\bibliographystyle{mnras}
\bibliography{bibliography}

\appendix

\section{X-ray vs $\gamma$-ray luminosity of SNRs}
\label{app:XvsG}
The luminosities of Galactic SNRs in the 1--100~GeV band in Figure~\ref{f:X-GammaSNR} are derived from the fluxes reported by \citet{aaa16}, except for Kepler's SNR (\citealt{xj21}, but see also \citealt{alb22}) and Tycho's SNRs \citep{gnb12}. Unabsorbed X-ray luminosities were derived from the fluxes reported in the Chandra SNR catalog (\url{https://hea-www.harvard.edu/ChandraSNR/snrcat_gal.html}), except for W28 \citep{rb02}, W30 \citep{fo94}, G20.0-00.2 \citep{ppg13}, W41 \citep{tll07}, W44 \citep{hss06}, W49B \citep{mdb06}, W51C \citep{kks95}, Cygnus Loop \citep{kkp84}, G78.2+2.1 \citep{lgr13}, HB21 \citep{la21}, CTB 109 \citep{rp97}, IC 443 \citep{tbr06}, Monoceros \citep{lns86}, Puppis A \citep{dlr13}, Vela Jr. \citep{tbt16}, MSH 11-62 \citep{str12}, PKS 1209-52 \citep{mlt88}, G298.6-0.0 \citep{bsn16},  MSH 15-56 \citep{tsc13}, RX J1713.7-3946 \citep{adl09}, CTB 37A \citep{sgh11}, G349.7+0.2 \citep{lsh05}, G357.7-00.1 \citep{gfs03}. 
To convert fluxes in luminosities, we adopt the distances presented in \citet{aaa16}, except for G020.0-00.2 and G024.7+00.6 (\citealt{gre19} and references therein), G298.6-00.0 \citep{bsn16}, Tycho's SNR \citep{rui04}, and Kepler's SNR \citep{srb16}.

\bsp	 
\label{lastpage}
\end{document}